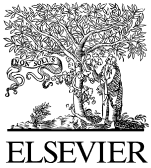

# New insights from cross-correlation studies between solar activity indices and cosmic-ray flux during Forbush decrease events

Mihailo Savić, Nikola Veselinović\*, Aleksandar Dragić, Dimitrije Maletić, Dejan Joković, Vladimir Udovičić, Radomir Banjanac, David Knežević

*Institute of Physics Belgrade, University of Belgrade, Pregrevica 118, 11080 Belgrade, Serbia*

**Abstract**

Observed galactic cosmic ray intensity can be subjected to a transient decrease. These so-called Forbush decreases are driven by coronal mass ejection induced shockwaves in the heliosphere. By combining in situ measurements by space borne instruments with ground-based cosmic ray observations, we investigate the relationship between solar energetic particle flux, various solar activity indices, and intensity measurements of cosmic rays during such an event. We present cross-correlation study done using proton flux data from the SOHO/ERNE instrument, as well as data collected during some of the strongest Forbush decreases over the last two completed solar cycles by the network of neutron monitor detectors and different solar observatories. We have demonstrated connection between the shape of solar energetic particles fluence spectra and selected coronal mass ejection and Forbush decrease parameters, indicating that power exponents used to model these fluence spectra could be valuable new parameters in similar analysis of mentioned phenomena. They appear to be better predictor variables of Forbush decrease magnitude in interplanetary magnetic field than coronal mass ejection velocities.
 2022 COSPAR. Published by Elsevier B.V. All rights reserved.

## 1. Introduction

Cosmic rays (CRs) are high-energy charged particles that arrive at Earth from space, mainly originating from outside of our Solar system. CRs are modulated in the heliosphere (Heber et al., 2006) due to interaction with the interplanetary magnetic field (IMF) frozen in a constant stream of charged particles from Sun - the solar wind (SW). Transients in the heliosphere additionally modulate CRs. One type of transients are interplanetary coronal mass ejections (ICMEs), closely related to coronal mass ejections (CMEs).

ICMEs interact with SW, and as the speed of particles in ICME is different than the speed of SW particles, a bow shock can be created, affecting the CR flux (Belov et al., 2014). This interaction between ICMEs and residual solar wind can be one of the causes of short-term depression in CR flux, detectable at Earth (Subramanian et al., 2009). Such transient decrease in observed flux is known as a Forbush decrease (FD), a type of CR flux modulation that has been studied extensively since its initial discovery in the 1930s (Gopalswamy (2016) and references therein). There are two clearly distinguishable classes of Forbush decreases: recurrent and non-recurrent. Non-recurrent FDs, typically caused by ICMEs (Dumbovic et al., 2012), are mostly characterized by a sudden offset, which lasts about a day, followed by a gradual recovery phase within several days (Cane, 2000). Due to ICME sub-structures (the sheath and the associated shock and magnetic cloud) FD can have one or two-step profile, which depends on transit of one or both structures to the observer (Richardson and Cane, 2011). Recurrent FDs have different profile, with gradual onset and decrease and symmetrical recovery caused by high-speed streams from coronal holes (Melkumyan et al., 2019). In this paper we will focus on non-recurrent ICME induced FDs.

Apart from FD profile, one of the main parameters that is used to describe a Forbush decrease is its magnitude. The effect is not the same for all CR particles, as it depends on their rigidity. Rigidity is defined as $R \equiv B\rho = p/q$, where $\rho$ is gyroradius of the particle due to magnetic field $B$, $p$ is particle momentum, and $q$ is its charge. The higher the rigidity of a particle, the less it is affected by heliospheric inhomogeneities, hence the reduction in flux is less pronounced.

Another phenomenon that can accompany violent events on the Sun is emission of fast-moving particles, commonly known as solar energetic particles (SEP). The occurrence of such particles is typically related to eruptions on the surface of the Sun, which can be characterized by bursts of X-rays - solar flares (SF), and/or emission of coronal plasma - already mentioned CMEs. When excess of these solar energetic particles with high energy penetrates the geomagnetic field, it can cause a sudden and brief increase in measured CR flux at Earth - a ground level enhancement (GLE). Because GLEs can be harmful to human infrastructures (potentially damaging power lines, satellites in orbit, etc.), they have been studied in detail for decades.

Variations of CR flux have been monitored at Earth for decades using ground and underground-based detectors, primarily neutron monitors (NM) (Belov et al., 2000; Koldobskiy et al., 2019) and muon detectors (Mendonça et al., 2016; Veselinović et al., 2015). Different types of ground-based detectors complement each other in terms of their CR energy domain (Veselinovic et al., 2017), muon detectors being sensitive to energies higher than those detectable by NMs. In addition, CR flux is also (especially in the last couple of decades) directly measured in space using space-borne instruments (Dumbovic et al., 2020; von Forstner et al., 2020). In the MeV energy range most space probe particle detectors are sensitive to, enhancement of SEP flux can enshroud CR flux, thus making a task of establishing decoupled event-integrated energy spectra (or spectral fluences) for SEP and CRs a laborious task (Koldobskiy et al., 2021; Bruno and Richardson, 2021).

Many authors have studied the connection between SFs, CMEs/ICMEs and SEP, consequential effects on the geomagnetic field and compound effect of the IMF and geomagnetic field disturbances on CRs. Most relevant for our analysis is work that studied connection between different FD and ICME parameters (Belov et al. (2000), Belov (2008), Papaioannou et al. (2020) and references therein), which has among other, shown significant correlation between CME speeds and FD magnitudes. More precisely, CME speeds have been established as the best predictor variables of FD magnitudes for primary CR particles with 10 GV rigidity detected at Earth. Also of interest is the work that studied the connection between the disturbance of geomagnetic field and CR flux measured at Earth (Alhassan et al., 2021; Badruddin et al., 2019), where a significant correlation between FD magnitude and different geomagnetic parameters due to common solar or interplanetary origin has been established.

SF, CME/ICME, SEP and FD events are very often related processes that occur either simultaneously or in succession, in which case can be thought of as different components of one more complex event. CMEs (along with their interplanetary counterparts ICMEs) have been recognized as the main driver of FDs, while on the other hand there has been plenty of evidence for the relationship between CMEs with SEP. Namely, there are two different known mechanism for SEP acceleration: acceleration during magnetic-reconnection events usually resulting in solar flares (which produce short impulsive SEP events), and acceleration caused by CME induced shock waves (which result in gradual SEP events) (Reames, 1999). For this study the second class is of interest. Another type of closely related events that are important for this analysis are energetic storm particle (ESP) events, which represent particles accelerated locally by interplanetary shocks driven by fast CMEs (Desai and Giacalone, 2016). Even though details of the mechanism and the precise role of CME induced shock in the evolution of SEP events are not fully understood (Anastasiadis et al., 2019), we believe that analysis of how SEP/ESP events relate to CME, geomagnetic and FD events could provide some valuable new insight. We are especially interested in, and will concentrate the most on, the possibility of the last of these connections. To do so, we have decided to look into the shape of SEP/ESP fluence spectra and analyze how it relates to different CME, geomagnetic and especially FD parameters.

It should be noted that different mentioned types of events, even when related, do not need to occur at the same place nor at the same time. This is due to the fact that SEP travel along magnetic field lines, while CME/ICME shocks travel mostly directly away from the Sun. Furthermore, modulation of primary CR, detected as FD upon their arrival at Earth, can happen anywhere in the heliosphere. Hence, in general case, detection of these events should not necessarily be simultaneous. However, we believe that for the class of events selected for this analysis we can assume that they occur and are detected within a certain time window. We will elaborate more on this in Section 2.3.

The article is structured as follows: first we list various sources of data and justify the selection of solar cycle 23 and 24 FD events to be used in the analysis; then we describe parametrization of SEP events (involving calculation and parametrization of SEP fluence spectra); finally we perform correlative analysis between established SEP parameters and various CME, FD and geomagnetic indices and discuss the observed dependencies.

## 2. Data

Sources of SEP proton flux, various solar and space weather parameters, as well as ground CR measurements and different FD parameters used in this study are listed below. Different criteria for FD event selection are also described.

### 2.1. Solar energetic particle flux data

The source for SEP flux data was the ERNE instrument (Torsti et al., 1995) onboard the Solar and Heliospheric Observatory (SOHO). Instrument consists of two separate particle detectors. The Low-Energy Detector (LED) and the High-Energy Detector (HED). Former covers ion fluxes and count rates in the $1.3 - 13$ MeV/nucleon energy range, and latter ion fluxes and count rates in the $13 - 130$ MeV/nucleon energy range. Both ranges are separated in ten energy channels. SOHO has been making in situ observation from Lagrangian point L1 for the last three solar cycles (data available at https://omniweb.gsfc.nasa.gov/ftpbrowser/flux_spectr_m.html). ERNE data for solar cycles 23, 24 and current cycle 25 allows the study of variations of proton fluences in SEP events during this period (Paassilta et al., 2017; Belov et al., 2021). Higher channels are more correlated with measured CR flux (Veselinovic et al., 2021) and it appears as if flux in these channels is a mixture of CR and energetic proton fluxes of particles with the same energy. Important feature of HED detector is that, due to rather large geometric factor, during large intensity proton events SOHO/ERNE data have been subject to saturation effects in higher energy channels (Valtonen and Lehtinen, 2009; Miteva et al., 2020).

### 2.2. IZMIRAN directory of Forbush decreases

IZMIRAN database is an online repository developed at the Institute of Terrestrial Magnetism, Ionosphere and Radiowave Propagation (IZMIRAN) at Moscow Troitsk, Russia. It contains an extensive list of Forbush decreases and various parameters from solar, space weather, cosmic ray and geomagnetic measurements, spanning from the late 1950s (http://spaceweather.izmiran.ru/eng/dbs.html). Database has been compiled from a number of sources, such as measurements by ground-based detectors, instruments mounted on various satellites, as well as public data provided by different agencies specializing in monitoring solar, space and atmospheric weather and geomagnetism. Extensive list of sources and data repositories used to compile this database are referenced in a number of publications listed on the IZMIRAN internet site (IZMIRAN Space Weather Prediction Center, 2016).

We have decided to use IZMIRAN database as our primary source of data for Forbush decrease parameters as well as for selected variables, parameters and indices that describe associated space weather and geomagnetic phenomena. Selection of parameters pertinent to our analysis was mostly based on previous work by other authors (i.e. Belov (2008), Lingri et al. (2016)), where they established which quantities are most relevant in these types of studies.

Chosen parameters fall into three categories (abbreviations to be used throughout the text are given in parentheses). First category are FD related parameters - Forbush decrease magnitude for 10 GV rigidity primary particles ($M$) and Forbush decrease magnitude for 10 GV rigidity primary particles corrected for magnetospheric effect using $Dst$ index ($M_M$). These magnitudes are determined using global survey method (GSM). GSM combines measurements from a world-wide network of neutron monitors (NMs), takes into account different anisotropies, disturbances of atmospheric and geomagnetic origin, as well as apparatus-specific features, and produces an estimated hourly variation of CR flux outside Earth's atmosphere and magnetosphere (Belov et al., 2018). Specifically, correction for magnetospheric effect takes into account the fact that geomagnetic disturbances affect the effective cutoff threshold rigidities and effective asymptotic directions of primary particles for different NM stations (Belov et al., 2005).

Second group of parameters used from IZMIRAN database are CME and SW related parameters - the average CME velocity between the Sun and the Earth, calculated using the time of the beginning of the associated X-ray flare ($V_{mean}$), the average CME velocity between the Sun and the Earth, calculated using the time of the beginning of the associated CME observations ($V_{meanC}$) and maximal hourly solar wind speed in the event ($V_{max}$). Izmiran DB authors have matched detected FD events with associated CMEs using a SOHO LASCO CME catalog (Belov et al., 2014). Catalog includes a comprehensive list of CME events along with some of most relevant parameters, i.e. speeds calculated by tracking CME leading edge (as described in Yashiro et al. (2004), further sources available at https://cdaw.gsfc.nasa.gov/CME_list/catalog_description.htm).

Final group of parameters from IZMIRAN database used in this analysis are related to geomagnetic field - maximal $Kp$ index in the event ($Kp_{max}$ - based on data from NOAA Space Weather Prediction Center, https://www.swpc.noaa.gov/products/planetary-k-index), maximal 3-h $Ap$ index in the event ($Ap_{max}$ - defined as the mean value of the variations of the terrestrial magnetic field, derived from Kp index) and minimal $Dst$ index in the event ($Dst_{min}$ - calculated using data provided by World Data Center for Geomagnetism, Kyoto, http://wdc.kugi.kyoto-u.ac.jp/dstdir/index.html).

### 2.3. Selection of FD events

Time interval used for this analysis was dictated by the period of operation of SOHO/ERNE device, which was commissioned in December 1995 (data available from June 1996) and is still operational. That coincides with the beginning of solar cycle 23 and lasts through cycle 24, so we considered all FD events that occurred in this period, concentrating on events with magnitudes for 10 GV particles larger 4% in the analysis. There are several reason for such magnitude cut, primary reason being that even though we often reference neutron monitor data in the analysis, CR related research in our laboratory is mainly based on muons detectors, which are generally less sensitive to FDs of smaller magnitude and GLE events. Additionally, it is known that all larger FDs (i.e. with magnitudes greater than 5%) are caused by CMEs (Belov, 2008). Since we use CME speed as a reference parameter in the analysis, introducing such cut made event selection simpler, as practically all considered FD events would have an associated CME. Finally, CME speed is less reliably determined in the case of weaker CME events (Yashiro et al., 2004).

One important step in the event selection procedure is to make sure that for each global event both proton flux increase detected by SOHO/ERNE and FD are related to the same CME. As mentioned in the introduction, detection of these separate events is not necessarily simultaneous. However, we have checked the direction of CMEs/ICMEs for all events for which such information was available, and in all these cases they moved directly toward Earth. This would imply that detection of the increase of energetic particles, Forbush decrease and geomagnetic storm associated with a given CME should be detectable within a relatively small time window. To illustrate this, on Fig. 1 we have shown time series for proton flux (in one selected energy channel), CR flux and $Dst$ index for one such event. Furthermore, because of large magnitudes of FDs selected for the analysis, we believe it to be the case for all events.

Another important point is that we cannot say with certainty what is the exact origin of detected proton flux solely based on SOHO/ERNE data. They could be of solar origin (SEP), particles accelerated locally at shock in interplanetary space (ESP), or combination of both. For the sake of simplicity we have decided to use the somewhat more general term SEP for these energetic particles, having mentioned limitation of its use in mind.

As determination of SEP fluence is not a straightforward procedure (as explained in more detail in Section 3.1), from the initial set of events we discarded all for which fluence value was difficult to determine or had a large uncertainty due to overlap and unclear separation of proton flux time series of successive events. That set was then further reduced based on the quality of FD identification flag assigned to each event in the IZMIRAN database, taking into account only events where identification was confident or reliable enough. Applying mentioned selection criteria resulted in the final set of 21 events, presented in Table 1 with some of the parameters of interest.

## 3. Parametrization of SEP fluence energy spectra

Parametrization procedure for any of the selected FD events can be broken down into two steps: 1 - calculation of SEP fluence in different energy channels and 2 - determination of power exponents for SEP fluence spectra.

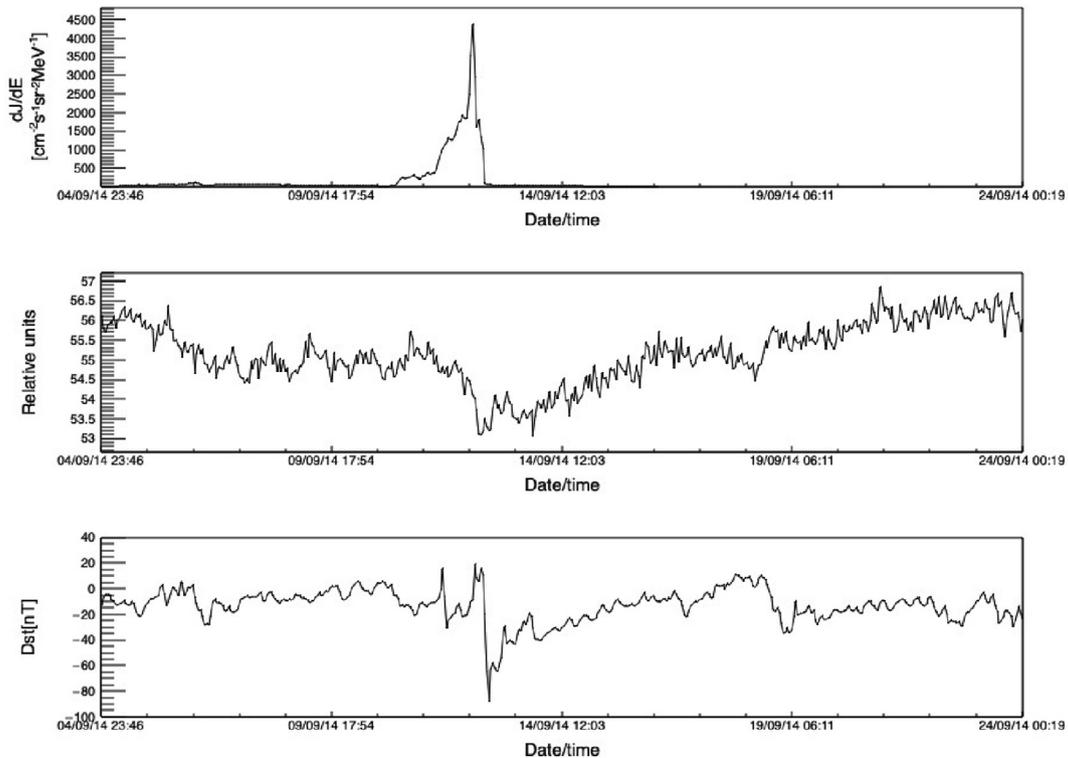

Fig. 1. Time series of hourly data for the same time interval around FD event of 12 Septemeber 2014: proton flux in the $1.3 - 1.6$ MeV channel (top), Athens neutron monitor count rate (middle), and Dst index (bottom).

Table 1
Forbush decrease events from solar cycles 23 and 24 selected for the analysis, along with some of the FD, CME and geomagnetic field parameters of interest.

| Date/Time | $M$ [%] | $M_M$ [%] | X flare | $V_{mean}$ [km s$^{-1}$] | $V_{meanC}$ [km s$^{-1}$] | $V_{max}$ [km s$^{-1}$] | $Kp_{max}$ | $Ap_{max}$ | $Dst_{min}$ [nT] |
|---|---|---|---|---|---|---|---|---|---|
| 2001.09.29 09:40:00 | 4.3 | 4.4 | M 1.0/ | 852.0 | 831 | 694.0 | 5.33 | 56.0 | −56.0 |
| 2001.10.11 17:01:00 | 7.0 | 6.9 | M 1.4/2F | 766.0 | 769 | 572.0 | 6.0 | 80.0 | −71.0 |
| 2001.10.21 16:48:00 | 5.4 | 7.3 | X 1.6/2B | 855.0 | 858 | 677.0 | 7.67 | 179.0 | −187.0 |
| 2001.11.24 05:56:00 | 9.2 | 9.8 | M 9.9/ | 1323.0 | 1366 | 1024.0 | 8.33 | 236.0 | −221.0 |
| 2002.04.17 11:07:00 | 6.2 | 7.0 | M 1.2/SF | 742.0 | 745 | 611.0 | 7.33 | 154.0 | −127.0 |
| 2002.09.07 16:36:00 | 4.6 | 5.1 | C 5.2/SF | 860.0 | 863 | 550.0 | 7.33 | 154.0 | −181.0 |
| 2003.10.30 16:19:00 | 14.3 | 9.4 | X10.0/2B | 2109.0 | 2140 | 1876.0 | 9.0 | 400.0 | −383.0 |
| 2003.11.20 08:03:00 | 4.7 | 6.8 | M 3.2/2N | 854.0 | 872 | 703.0 | 8.67 | 300.0 | −422.0 |
| 2004.07.26 22:49:00 | 13.5 | 14.4 | M 1.1/1F | 1279.0 | 1290 | 1053.0 | 8.67 | 300.0 | −197.0 |
| 2004.09.13 20:03:00 | 5.0 | 5.3 | M 4.8/SX | 945.0 | 948 | 613.0 | 5.33 | 56.0 | −50.0 |
| 2005.05.15 02:38:00 | 9.5 | 12.2 | M 8.0/SX | 1207.0 | 1231 | 987.0 | 8.33 | 236.0 | −263.0 |
| 2006.12.14 14:14:00 | 8.6 | 9.6 | X3.4/4B | 1154.0 | 1165 | 955.0 | 8.33 | 236.0 | −146.0 |
| 2011.02.18 01:30:00 | 5.2 | 4.7 | X2.2/ | 579.0 | 579 | 691.0 | 5.0 | 48.0 | −30.0 |
| 2011.08.05 17:51:00 | 4.3 | 4.8 | M 9.3/ | 1089.0 | 1104 | 611.0 | 7.67 | 179.0 | −115.0 |
| 2011.10.24 18:31:00 | 4.9 | 6.5 | - | - | 633 | 516.0 | 7.33 | 154.0 | −147.0 |
| 2012.03.08 11:03:00 | 11.7 | 11.2 | X5.4/ | 1187.0 | 1188 | 737.0 | 8.0 | 207.0 | −143.0 |
| 2012.07.14 18:09:00 | 6.4 | 7.6 | X 1.4/ | 822.0 | 834 | 667.0 | 7.0 | 132.0 | −127.0 |
| 2013.06.23 04:26:00 | 5.9 | 5.3 | M 2.9/ | 832.0 | 844 | 697.0 | 4.33 | 32.0 | −49.0 |
| 2014.09.12 15:53:00 | 8.5 | 5.9 | X1.6/2B | 893.0 | 897 | 730.0 | 6.33 | 94.0 | −75.0 |
| 2015.06.22 18:33:00 | 8.4 | 9.1 | M2.6/ | 1027.0 | 1040 | 742.0 | 8.33 | 236.0 | −204.0 |
| 2017.09.07 23:00:00 | 6.9 | 7.7 | X9.3/ | - | 1190 | 817.0 | 8.33 | 236.0 | −124.0 |

### 3.1. SEP fluence calculation

SEP fluence is calculated by integrating SOHO/ERNE proton flux time series in separate energy channels over time period associated with a given FD event. First step in this procedure is to determine this time period (and hence integration boundaries) as precisely as possible. Most more energetic events we considered for this analysis have a strong SF associated with them. This may lead to a complex picture, as FD event of interest often occurs in the middle of a turbulent period where additional FDs (sometimes associated with other CMEs) precede or follow it. As a consequence, clear separation of successive events and determination of optimal integration boundaries may not be simple nor straightforward. To make this procedure more reliable, we have used IZMIRAN database and neutron monitor data (courtesy of the Neutron Monitor Database (Neutron Monitor Database, 2022)) in parallel with SOHO/ERNE proton time series, trying to identify prominent features in all three sources, so we could separate events of interest in all energy channels as clearly as possible.

Baseline for integration was determined based on a data interval of at least one (but preferably several) days, where proton flux was negligibly different from zero relative to the flux during the event. If possible, time interval before the event was taken for the calculation of baseline unless there was a preceding disturbance, in which case quiet interval following the event was taken instead. Integration of fluence for several selected SOHO/ERNE energy channels for the event of 12 September 2014 is shown on Fig. 2. Integration interval is indicated with vertical dashed lines and baseline value with a horizontal dashed line.

One interesting feature that can be observed in SOHO/ERNE data time series is that in some cases proton flux in the highest energy channels can dip below the baseline after the initial increase. For a number of events such behavior is even more pronounced, where in extreme cases it can happen that no flux increase is observed, but rather just the decrease. We believe this indicates that the highest energy channels have non-negligible contribution of low-energy cosmic rays, which can increase uncertainty for fluence calculation. We will refer to this again when discussing fluence spectra in Section 3.2.

To make fluence calculation procedure more reliable we have assigned a quality flag to each event, based on our estimate of the uncertainty of integration, and decided on a quality cut we deemed acceptable for further analysis. As mentioned in Section 2.3, 21 events have passed this criterium. Even then, for a number of events calculated fluence proved to be sensitive to small variations of integration boundaries, which makes it especially difficult to give a reliable estimate of the error for the integration procedure and should be kept in mind when discussing the results.

### 3.2. Determination of SEP fluence spectra power exponents

Fluence energy spectra for all selected events were formed using values for different energy channels, calculated as explained in the previous section. The choice of parameters to be used to describe their shape and characteristics depends on the analytic expression used to model the spectrum. In general, during a SEP event spectra exhibit a characteristic "bend" or a "knee", which is not so straightforward to describe theoretically. Various expressions were proposed to model this observed feature (Ellison and Ramaty, 1985; Mottl et al., 2001), out of which we have decided to use the following double power law one (Band et al., 1993; Zhao et al., 2016), as we feel it is well suited for our analysis:

$$\frac{dJ}{dE} = \begin{cases} E^{-\alpha} \exp\left(-\frac{E}{E_b}\right) & E \leqslant (\beta - \alpha)E_b, \\ E^{-\beta}[(\beta - \alpha)E_b]^{\beta - \alpha} \exp(\alpha - \beta) & E > (\beta - \alpha)E_b, \end{cases}$$
(1)

where $E_b$ is knee energy at which the break occurs, while $\alpha$ and $\beta$ are power-law exponents that describe energy ranges below and above the break respectively, and consequently are variables we chose to parametrize the SEP event.

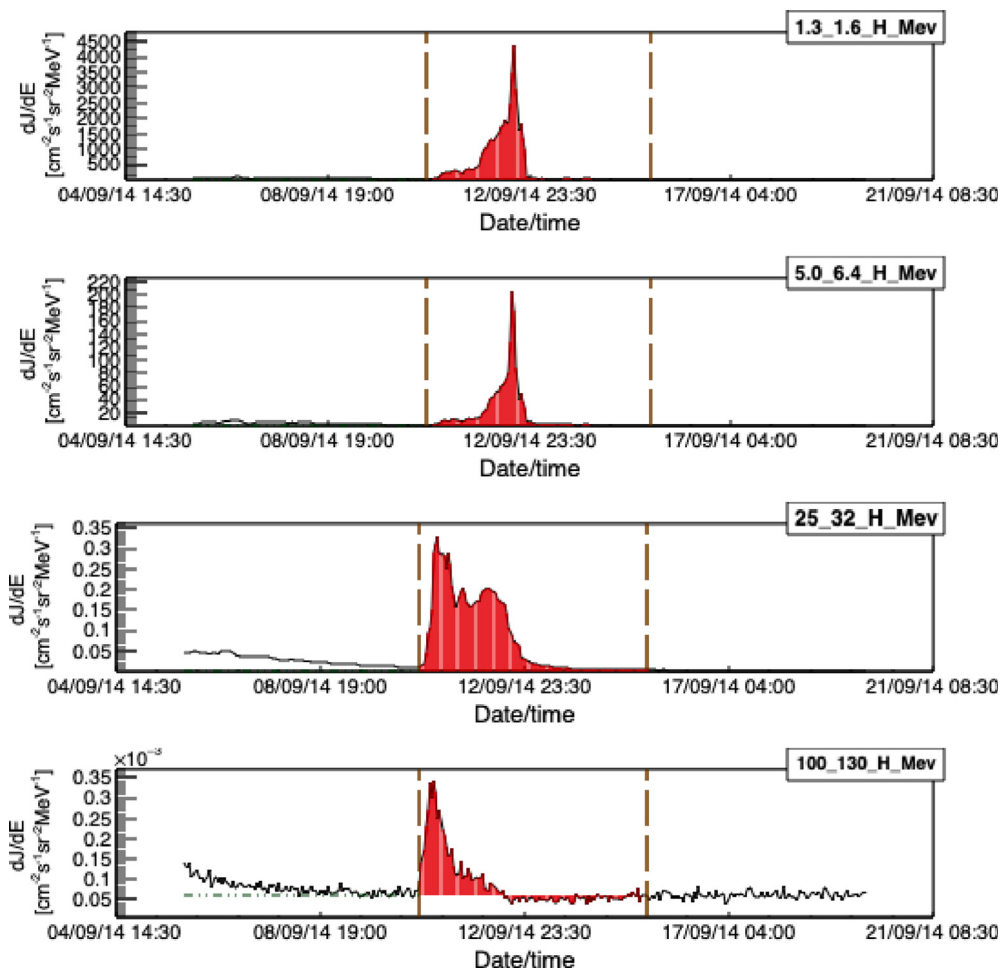

Fig. 2. Solar proton flux for four selected energy channels during FD event of 12 September 2014. Vertical dashed lines indicate integration interval, horizontal dashed line indicates the baseline value, while areas shaded red correspond to result of the integration used to calculate the SEP fluence.

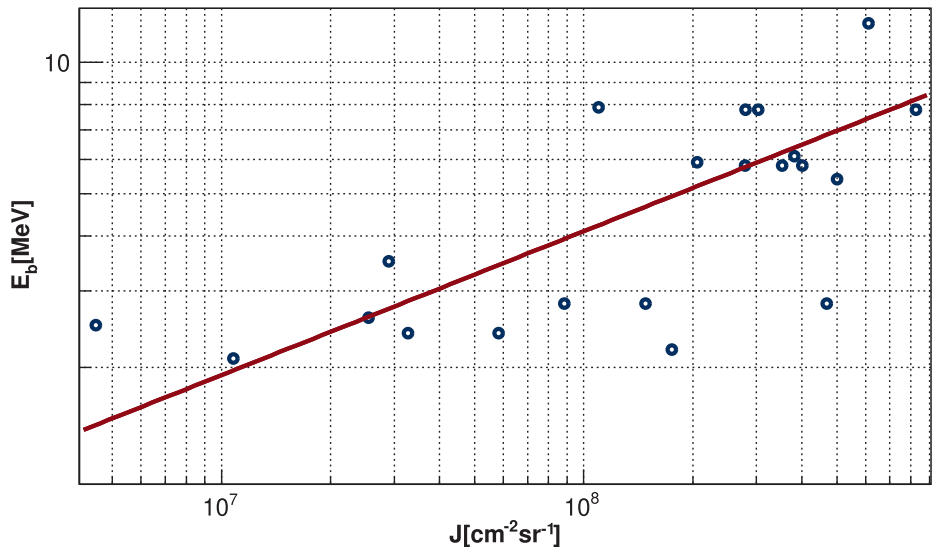

Fig. 3. "Knee" energy dependence on SEP fluence (integrated over full energy range) for selected events. Power function fit is indicated by the red line.

These power-law exponents obtained by fitting fluence spectra with Expression 1 can be very sensitive to variation of knee energy, so some care needs to be taken in order to determine $E_b$ as accurately as possible.

Determination of knee energy using "by eye" method proved to be uncertain enough for us to decide on using a more quantitative approach, which is based on the fact that knee energy generally depends on the integral fluence of the event (as described in Nymmik (2013) and Miroshnichenko and Nymmik (2014)). In accordance with this, we firstly determined the knee energy "by eye", plotted it against integral fluence and then fitted this dependence with a power function in the form of $E_b = aJ^b$ (Fig. 3), where $E_b$ is the knee energy, $J$ integral fluence, and $a$ and $b$ are fit parameters. We then used these fit parameteres to determine $E_b$ for each event. In several cases where there has been some overlap of proton flux time series profiles associated with different successive events, small correction for integral fluence was introduced, which also affected the knee energy value.

Fluence spectra were then fitted with expression given in Eq. 1, using thusly calculated knee energy. On Fig. 4 we can see two characteristic examples that illustrate how well this expression actually models the fluence spectrum during a SEP event. In case of 11 October 2001 event (Fig. 4a) we see that the theoretical model fits the experimental data reasonably well, except for some small disagreement in the highest energy channels (feature we believe can be explained by our assumption that there is a non-negligible contribution of low-energy CR in this energy range). On the other hand, for a number of events with greater SEP flux higher energy channels tend to get saturated (as mentioned in Section 2.1). This in turn leads to an underestimated fluence and consequently poorer fit in this energy range, as can be seen for the 24 November 2001 event shown on Fig. 4b. Contribution of flux in these high-energy channels to integral fluence is very small, so this underestimated value does not significantly affect the value of knee energy or uncertainty of the exponent $\alpha$. However, the uncertainty of exponent $\beta$ is more significantly affected and for this reason in further analysis we will rely on exponent $\alpha$ more for the parametrization of fluence spectra.

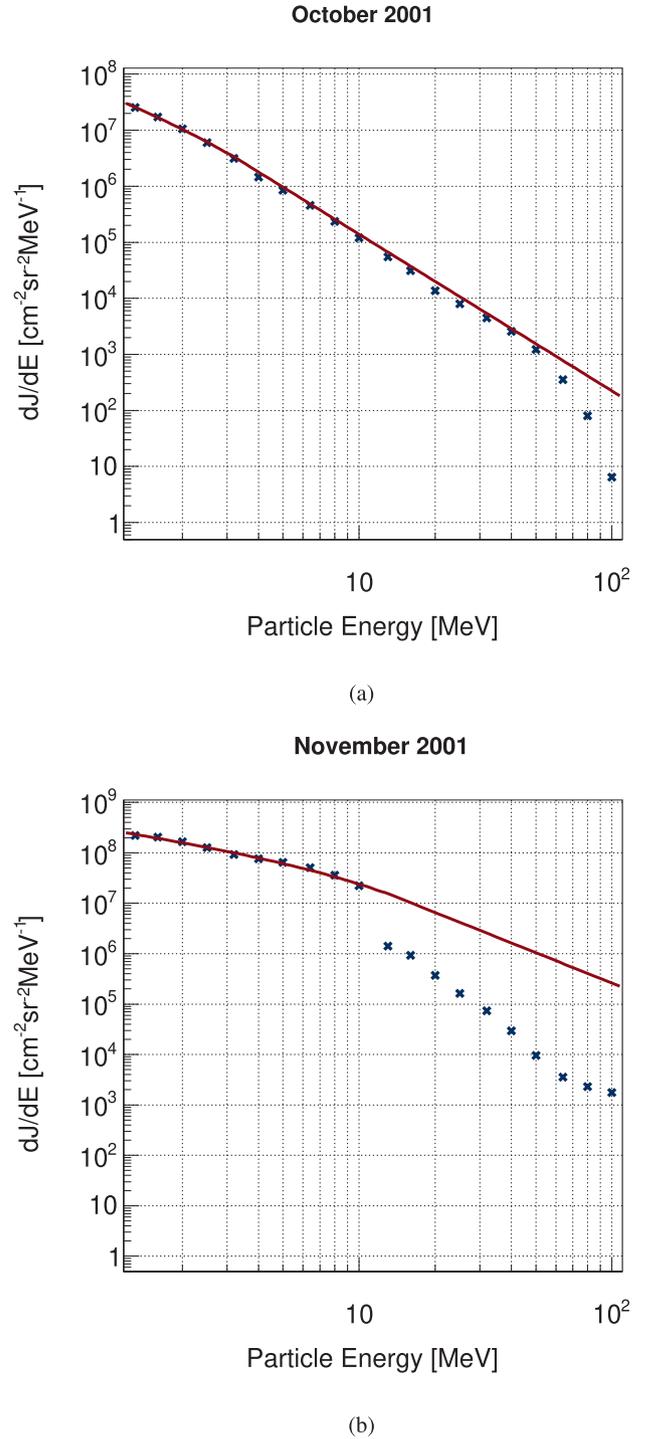

Fig. 4. SEP fluence energy spectra for the: (a) 11 October 2001 event, (b) 24 November 2001 event. Red lines indicate the double power law fit.

## 4. Correlative analysis

We have performed correlative analysis between power exponents chosen to parametrize SEP fluence and selected parameters from Izmiran database. The results are presented in Table 2. Worth noting is the slightly lower statistics for $V_{mean}$ due to exclusion of two events for which this parameter was not available.

Table 2
Correlation coefficients (r) between SEP fluence spectra power exponents and selected FD, CME and geomagnetic field indices.

|  | $\alpha$ | $\beta$ | $M$ | $M_M$ | $V_{meanC}$ | $V_{mean}$ | $V_{max}$ | $Kp_{max}$ | $Ap_{max}$ | $Dst_{min}$ |
|---|---|---|---|---|---|---|---|---|---|---|
| $\alpha$ | 1.00 | 0.96 | 0.67 | 0.64 | 0.77 | 0.75 | 0.66 | 0.40 | 0.53 | −0.40 |
| $\beta$ | 0.96 | 1.00 | 0.67 | 0.67 | 0.72 | 0.70 | 0.60 | 0.44 | 0.50 | −0.38 |
| $M$ | 0.67 | 0.67 | 1.00 | 0.84 | 0.79 | 0.79 | 0.79 | 0.53 | 0.65 | −0.41 |
| $M_M$ | 0.64 | 0.67 | 0.84 | 1.00 | 0.57 | 0.57 | 0.53 | 0.69 | 0.69 | −0.46 |
| $V_{meanC}$ | 0.77 | 0.72 | 0.79 | 0.57 | 1.00 | 1.00 | 0.92 | 0.61 | 0.77 | −0.58 |
| $V_{mean}$ | 0.75 | 0.70 | 0.79 | 0.57 | 1.00 | 1.00 | 0.92 | 0.62 | 0.78 | −0.60 |
| $V_{max}$ | 0.66 | 0.60 | 0.79 | 0.53 | 0.92 | 0.92 | 1.00 | 0.49 | 0.71 | −0.58 |
| $Kp_{max}$ | 0.40 | 0.44 | 0.53 | 0.69 | 0.61 | 0.62 | 0.49 | 1.00 | 0.94 | −0.78 |
| $Ap_{max}$ | 0.53 | 0.50 | 0.65 | 0.69 | 0.77 | 0.78 | 0.71 | 0.94 | 1.00 | −0.87 |
| $Dst_{min}$ | −0.40 | −0.38 | −0.41 | −0.46 | −0.58 | −0.60 | −0.58 | −0.78 | −0.87 | 1.00 |

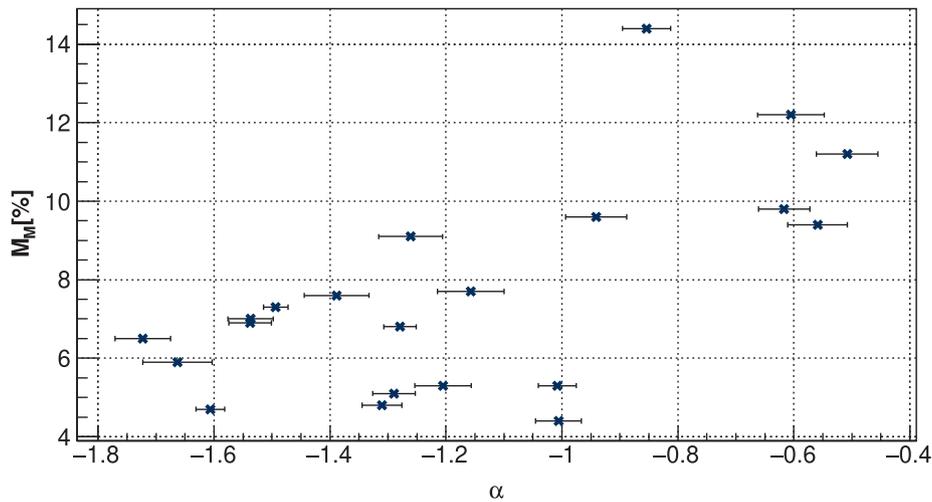

Fig. 5. Dependence of FD magnitude for particles with 10 GV rigidity corrected for magnetospheric effects ($M_M$) on power exponent $\alpha$.

Strong correlation between FD magnitude for particles with 10 GV rigidity ($M$) and mean CME ($V_{meanC}$, $V_{mean}$) and maximum SW ($V_{max}$) velocities illustrates the important role these parameters have in driving FD events, as has been discussed in detail by several authors (i.e. Belov et al. (2014)). On the other hand, correlation between these velocities and parameter $M_M$ is noticeably smaller. $M_M$ is FD magnitude for particles with 10 GV rigidity corrected for magnetospheric effect (using $Dst$ index), so we could approximate it as an estimated measure of the FD magnitude in interplanetary magnetic field.

If we now look at how SEP fluence spectra power exponents relate to other parameters in Table 2, we observe the best correlation with mean CME velocities, while it is somewhat smaller with maximum SW velocity. Correlation with FD magnitude ($M$) is smaller than for CME velocities, however interestingly the correlation with the corrected FD magnitude ($M_M$) appears larger than in the case of CME velocities. One possible explanation for this could be that the shape of SEP fluence spectrum is more related to CR disturbance induced in interplanetary magnetic and less to one induced in geomagnetic field. What could support this assumption further is the fact that we observe smaller correlation between $\alpha$ and $\beta$ exponents and geomagnetic indices $Kp_{max}$, $Ap_{max}$ and $Dst_{min}$ than between these indices and CME velocities.

It should be said that even though SEP fluence spectra power exponents are not directly measured independent variables, the procedure to calculate them is relatively simple, while procedure used to calculate FD magnitudes (using GSM approach) is somewhat less straightforward and accessible. Hence, these exponents could be used to give a first estimate of Forbush decrease magnitudes outside atmosphere and magnetosphere. Having this in mind, we could conclude that SEP fluence power exponents could be better predictor variables (in the sense described above) of FD magnitude in interplanetary space than CME velocities are, while they are less reliable predictor variables of FD magnitude observed at Earth. If true, this could possibly lead us a small step closer to empirically decoupling the effects of IMF and geomagnetic fields on CR.

To further examine how FD magnitude corrected for magnetospheric effects is related to the shape of SEP fluence spectra, we have analyzed their dependence, which is plotted on Fig. 5. Both power exponents exhibit similar dependence, but only plot for $\alpha$ is shown, as it has consid-erably smaller uncertainty (as mentioned in Section 3.2) and we believe it to be a more reliable parameter. We can see that the graph is fairly linear, as could be expected based on the correlation coefficients, but on closer inspection it appears as if there are two separate classes of events with somewhat different behavior. If we loosely divide all FD events into low-magnitude set (with $M_M$ less than 6%) and high-magnitude set (with $M_M$ greater or equal to 6%), we can observe much weaker dependence of corrected FD magnitude on power exponent α for the first class than for the second one.

Table 3
Correlation coefficients (r) between FD magnitudes for particles with 10 GV rigidity (uncorrected $M$ and corrected for magnetospheric effect $M_M$) and SEP fluence spectra power exponents, selected FD, CME and geomagnetic field indices for particles with $M_M \geqslant 6\%$ (left) and particles with $M_M < 6\%$ (right).

|       | $M_M \geqslant 6\%$ | | | | | $M_M < 6\%$ | | | | |
|-------|---|---|---|---|---|---|---|---|---|---|
|       | $\alpha$ | $\beta$ | $V_{meanC}$ | $V_{mean}$ | $V_{max}$ | $\alpha$ | $\beta$ | $V_{meanC}$ | $V_{mean}$ | $V_{max}$ |
| $M$   | 0.82 | 0.76 | 0.84 | 0.85 | 0.78 | −0.55 | −0.25 | −0.08 | −0.10 | 0.62 |
| $M_M$ | 0.77 | 0.76 | 0.52 | 0.49 | 0.55 | −0.38 | 0.01 | 0.23 | 0.19 | 0.17 |

To check if this observation is well founded, we look into the correlation coefficients for these two separate classes, which are shown in Table 3.

We can see that correlation coefficients for these two sets are indeed very different. While in case of FDs with $M_M$ equal or greater than 6% we observe an even larger correlation than before between power exponents α and β and both FD magnitude and corrected FD magnitude (approaching the values of correlation coefficients for CME velocities), coefficients for FDs with $M_M$ less than 6% have very different values, correlation even being negative. Although statistics for this second set of events is rather small (and hence the uncertainty for correlation coefficients might be large), it appears that the assumption about two classes of events does stand. What is more, we observe a similarly drastic difference in correlation coefficients between FD magnitudes and mean CME velocities (with little to none correlation for events with $M_M < 6\%$), also pointing to the existence of two separate classes of events. This could need to be further confirmed using larger statistics, i.e. by including FD events with magnitudes smaller than 4%.

## 5. Conclusions

We analyzed the connection between CME, SEP and FD events, investigating how the shape of SEP fluence spectra during the global disturbance relates to different CME and FD parameters typically used in such analysis. We fitted SEP fluence spectra with double power law and used power exponents (α and β) from these fits to parametrize the shape of SEP fluence spectra.

By the means of correlative analysis we investigated the connection between SEP fluence spectra power exponents and selected CME and SW parameters (mean CME and maximum SW velocities), as well as selected FD parameters (magnitude for 10 GV particles and magnitude for 10 GV particles corrected for magnetospheric effect) and various parameters of geomagnetic activity ($Kp, Ap$ and $Dst$ indices).

We observed largest correlation between power exponents and CME velocities. The correlation between power exponents and FD magnitude ($M$) is significant yet smaller than in case of mean CME velocities ($V_{meanC}, V_{mean}$) and FD magnitude. On the other hand, the correlation between FD magnitude corrected for magnetospheric effects ($M_M$) and power exponents is larger than between these magnitudes and mean CME velocities.

The dependence of corrected FD magnitude on power exponent α possibly indicates two separate classes of events in terms of corrected magnitude value, rough boundary being corrected FD magnitude value of 6%. Events with corrected FD magnitude larger than 6% show increased correlation with power exponent α, while for the set of events with this magnitude smaller than 6% correlation even has opposite sign. Similarly considerable difference between two classes of events can be observed in correlations of mean CME velocities and corrected FD magnitude. Even taking into account smaller number of events used in the analysis, this could be an indication of these two groups of events exhibiting different behavior.

With everything considered, we believe we have demonstrated an important connection of the shape of SEP fluence spectra with CME and FD events, and that power exponents α and β can be valuable new parameters to be used in the future study of mentioned phenomena. They seem to be better predictor variables of FD magnitude (and hence CR disturbance) in interplanetary magnetic field than CME velocities, especially in the case of events where FD magnitude corrected for magnetospheric effect is larger than 6%.

**Declaration of Competing Interest**

The authors declare that they have no known competing financial interests or personal relationships that could have appeared to influence the work reported in this paper.

**Acknowledgments**

The authors acknowledge funding provided by the Institute of Physics Belgrade, through the grant by the Ministry of Education, Science and Technological Development of the Republic of Serbia.

OMNI data was made available by NASA/GSFC's Space Physics Data Facility's OMNIWeb service. Data from the SOHO experiment, an international collaboration between ESA and NASA, was kindly provided by ERNE team from Turku University, Finland. Neutron monitor data is available online through the use of excellent NEST tool, provided by the Neutron Monitor Database. We would also like to express our gratitude to the cosmic ray group at the IZMIRAN Space Weather Prediction Center at Pushkov Institute of Terrestrial Magnetism, Ionosphere and Radio Wave Propagation of the Russian Academy of Sciences for kindly providing catalogue of Forbush-effects and interplanetary disturbances.

Finally, we would like to thank the Reviewers for constructive comments and useful suggestions that significantly contributed to the quality of the manuscript.

## References

Alhassan, J.A., Okike, O., Chukwude, A.E., 2021. Testing the effect of solar wind parameters and geomagnetic storm indices on galactic cosmic ray flux variation with automatically-selected forbush decreases. Res. Astron. Astrophys. 21 (9), 234. https://doi.org/10.1088/1674-4527/21/9/234.

Anastasiadis, A., Lario, D., Papaioannou, A., et al., 2019. Solar energetic particles in the inner heliosphere: status and open questions. Philosoph. Trans. Roy. Soc. A: Mathe. Phys. Eng. Sci. 377 (2148), 20180100. https://doi.org/10.1098/rsta.2018.0100, URL: https://royalsocietypublishing.org/doi/abs/10.1098/rsta.2018.0100.

Badruddin, B., Aslam, O.P. M., Derouich, M. et al., 2019. Forbush decreases and geomagnetic storms during a highly disturbed solar and interplanetary period, 4–10 september 2017. Space Weather, 17(3), 487–496. URL: https://agupubs.onlinelibrary.wiley.com/doi/abs/10.1029/2018SW001941. https://doi.org/10.1029/2018SW001941.

Band, D., Matteson, J., Ford, L., et al., 1993. BATSE Observations of Gamma-Ray Burst Spectra. I. Spectral Diversity. Astrophys. J. 413, 281–292. https://doi.org/10.1086/172995.

Belov, A., 2008. Forbush effects and their connection with solar, interplanetary and geomagnetic phenomena. Proc. Int. Astron. Union 4 (S257), 439–450. https://doi.org/10.1017/S1743921309029676.

Belov, A., Abunin, A., Abunina, M., et al., 2014. Coronal mass ejections and non-recurrent forbush decreases. Sol. Phys. 289, 3949–3960. https://doi.org/10.1007/s11207-014-0534-6.

Belov, A., Baisultanova, L., Eroshenko, E., et al., 2005. Magnetospheric effects in cosmic rays during the unique magnetic storm on november 2003. J. Geophys. Res.: Space Phys. 110 (A09S20). https://doi.org/10.1029/2005JA011067.

Belov, A., Eroshenko, E., Oleneva, V., et al., 2000. What determines the magnitude of forbush decreases? Adv. Space Res. 27 (3), 625–630. https://doi.org/10.1016/S0273-1177(01)00095-3, URL: https://www.sciencedirect.com/science/article/pii/S0273117701000953.

Belov, A., Eroshenko, E., Yanke, V., et al., 2018. The Global Survey Method Applied to Ground-level Cosmic Ray Measurements. Sol. Phys. 293 (4), 68. https://doi.org/10.1007/s11207-018-1277-6.

Belov, A., Papaioannou, A., Abunina, M., et al., 2021. On the rigidity spectrum of cosmic-ray variations within propagating interplanetary disturbances: Neutron monitor and SOHO/EPHIN observations at ∼1-10 GV. Astrophys. J. 908 (1), 5. https://doi.org/10.3847/1538-4357/abd724.

Bruno, A., Richardson, I.G., 2021. Empirical model of 10–130 mev solar energetic particle spectra at 1 au based on coronal mass ejection speed and direction. Sol. Phys. 296 (36). https://doi.org/10.1007/s11207-021-01779-4.

Cane, H., 2000. Coronal mass ejections and forbush decreases. Space Sci. Rev. 93 (1), 55–77. https://doi.org/10.1023/A:1026532125747.

Desai, M., Giacalone, J., 2016. Large gradual solar energetic particle events. Living Rev. Sol. Phys. 13 (3). https://doi.org/10.1007/s41116-016-0002-5.

Dumbovic, M., Vršnak, B., Calogovic, J., et al., 2012. Cosmic ray modulation by different types of solar wind disturbances. A&A 538, A28. https://doi.org/10.1051/0004-6361/201117710.

Dumbovic, M., Vrsnak, B., Guo, J., et al., 2020. Evolution of coronal mass ejections and the corresponding forbush decreases: Modeling vs. multi-spacecraft observations. Sol. Phys. 295 (104). https://doi.org/10.1007/s11207-020-01671-7.

Ellison, D.C., Ramaty, R., 1985. Shock acceleration of electrons and ions in solar flares. Astrophys. J. 298, 400–408. https://doi.org/10.1086/163623.

Freiherr von Forstner, J.L., Guo, J., Wimmer-Schweingruber, R.F., et al., 2020. Comparing the properties of icme-induced forbush decreases at earth and mars. J. Geophys. Res.: Space Phys. 125(3), e2019JA027662. URL: https://agupubs.onlinelibrary.wiley.com/doi/abs/10.1029/2019JA027662. https://doi.org/10.1029/2019JA027662. E2019JA027662 10.1029/2019JA027662.

Gopalswamy, N., 2016. History and development of coronal mass ejections as a key player in solar terrestrial relationship. Geosci. Lett. 3 (8), 18. https://doi.org/10.1186/s40562-016-0039-2.

Heber, B., Fichtner, H., Scherer, K., 2006. Solar and heliospheric modulation of galactic cosmic rays. Space Sci. Rev. 125 (1), 81–91. https://doi.org/10.1007/s11214-006-9048-3.

IZMIRAN Space Weather Prediction Center, 2016. Izmiran space weather prediction center. URL: http://spaceweather.izmiran.ru/eng/about.html [Online; accessed 29-January-2022].

Koldobskiy, S.A., Bindi, V., Corti, C., et al., 2019. Validation of the neutron monitor yield function using data from ams-02 experiment, 2011–2017. J. Geophys. Res.: Space Phys. 124 (4), 2367–2379. https://doi.org/10.1029/2018JA026340, URL: https://agupubs.onlinelibrary.wiley.com/doi/abs/10.1029/2018JA026340.

Koldobskiy, S., Raukunen, O., Vainio, R., et al., 2021. New reconstruction of event-integrated spectra (spectral fluences) for major solar energetic particle events. Astron. Astrophys. 647, A132. https://doi.org/10.1051/0004-6361/202040058.

Lingri, D., Mavromichalaki, H., Belov, A., et al., 2016. Solar activity parameters and associated forbush decreases during the minimum between cycles 23 and 24 and the ascending phase of cycle 24. Sol. Phys. 291, 1025–1041. https://doi.org/10.1007/s11207-016-0863-8.

Melkumyan, A., Belov, A., Abunina, M., et al., 2019. On recurrent Forbush Decreases. In: Lagutin, A., Moskalenko, I., Panasyuk, M. (Eds.), Journal of Physics Conference Series, IOP Publishing, Bristol, United Kingdom volume 1181 of Journal of Physics Conference Series. p. 012009, https://doi.org/10.1088/1742-6596/1181/1/012009.

de Mendonça, R.R. S., Braga, C.R., Echer, E., et al., 2016. The temperature effect in secondary cosmic rays (Muons) observed at the ground: analysis of the global muon detector network data. Astrophys. J. 830(2), 88. https://doi.org/10.3847/0004-637x/830/2/88.

Miroshnichenko, L., Nymmik, R., 2014. Extreme fluxes in solar energetic particle events: Methodological and physical limitations. Radiation Measur. 61, 6–15. https://doi.org/10.1016/j.radmeas.2013.11.010, URL: https://www.sciencedirect.com/science/article/pii/S1350448713003806.

Miteva, R., Samwel, S.W., Zabunov, S., et al., 2020. On the flux saturation of SOHO/ERNE proton events. Bulgarian Astron. J. 33, 99.

Mottl, D.A., Nymmik, R. A., Sladkova, A.I., 2001. Energy spectra of high-energy SEP event protons derived from statistical analysis of experimental data on a large set of events. In: El-Genk, M.S., Bragg, M.J. (Eds.), Space Technology and Applications International Forum - 2001, AIP Publishing LLC., New York volume 552 of American Institute of Physics Conference Series, pp. 1191–1196, https://doi.org/10.1063/1.1358071.

Neutron Monitor Database, 2022. Neutron Monitor Database. URL: https://www.nmdb.eu/.

Nymmik, R., 2013. Charge states of heavy ions, as determined from the parameters of solar energetic particle spectra. Bull. Russian Acad. Sci.: Phys. 77, 490–492. https://doi.org/10.3103/S1062873813050419.

Paassilta, Miikka, Raukunen, Osku, Vainio, Rami, et al., 2017. Catalogue of 55–80 mev solar proton events extending through solar cycles 23 and 24. J. Space Weather Space Clim. 7, A14. https://doi.org/10.1051/swsc/2017013.

Papaioannou, A., Belov, A., Abunina, M., et al., 2020. Interplanetary coronal mass ejections as the driver of non-recurrent forbush decreases. Astrophys. J. 890 (2), 101. https://doi.org/10.3847/1538-4357/ab6bd1.

Reames, D.V., 1999. Particle acceleration at the sun and in the heliosphere. Space Sci. Rev. 90 (3), 413–491. https://doi.org/10.1023/A:1005105831781.

Richardson, I.G., Cane, H.V., 2011. Galactic Cosmic Ray Intensity Response to Interplanetary Coronal Mass Ejections/Magnetic Clouds in 1995–2009. Sol. Phys. 270 (2), 609–627. https://doi.org/10.1007/s11207-011-9774-x.

Subramanian, P., Antia, H.M., Dugad, S.R., et al., 2009. Forbush decreases and turbulence levels at coronal mass ejection fronts. A&A 494 (3), 1107–1118. https://doi.org/10.1051/0004-6361:200809551.

Torsti, J., Valtonen, E., Lumme, M., et al., 1995. Energetic particle experiment erne. Sol. Phys. 162 (1–2), 505–531. https://doi.org/10.1007/BF00733438.

Valtonen, E., Lehtinen, I.-V., 2009. Solar energetic particle fluences from soho/erne. Acta Geophys. 57, 116–124. https://doi.org/10.2478/s11600-008-0056-4.

Veselinovic, N., Dragic, A., Savic, M., et al., 2017. An underground laboratory as a facility for studies of cosmic-ray solar modulation. Nucl. Instrum. Methods Phys. Res. Section A: Accelerat. Spectromet. Detectors Assoc. Equip. 875, 10–15. URL: https://www.sciencedirect.com/science/article/pii/S0168900217309634. https://doi.org/10.1016/j.nima.2017.09.008.

Veselinović, Nikola, Savic, Mihailo, Dragic, Aleksandar, et al., 2021. Correlation analysis of solar energetic particles and secondary cosmic ray flux. Eur. Phys. J. D 75 (6), 173. https://doi.org/10.1140/epjd/s10053-021-00172-x.

Veselinović, N., Dragić, A., Maletić, D., et al., 2015. Cosmic rays muon flux measurements at Belgrade shallow underground laboratory. In: Trache, L., Chesneanu, D., Alexandru Ur, C. (Eds.), Exotic Nuclei and Nuclear/Particle Astrophysics (V) From Nuclei to Stars: Carpathian Summer School of Physics 2014, AIP Publishing LLC., New York volume 1645 of American Institute of Physics Conference Series, pp. 421–425. https://doi.org/10.1063/1.4909614.

Yashiro, S., Gopalswamy, N., Michalek, G., et al., 2004. A catalog of white light coronal mass ejections observed by the soho spacecraft. J. Geophys. Res.: Space Phys. 109 (A7). https://doi.org/10.1029/2003JA010282.

Zhao, L., Zhang, M., Rassoul, H.K., 2016. Double power laws in the event-integrated solar energetic particle spectrum. Astrophys. J. 821 (1), 62. https://doi.org/10.3847/0004-637x/821/1/62.